\documentclass[twocolumn]{aastex62}
\usepackage{import}

\usepackage{mathtools}
\usepackage{bm}
\usepackage{amsmath}
\usepackage{amssymb}
\usepackage{chngcntr}
\usepackage{footnote}
\usepackage[caption=false]{subfig}
\usepackage{float}
\usepackage{tabulary,booktabs}
\usepackage{multirow}

\shorttitle{MZR of a protocluster at $z \sim 2.2$}
\shortauthors{Sattari et al.}

\begin{document}

\title{\large \textbf{Evidence for gas-phase metal deficiency in massive protocluster galaxies at $z \sim 2.2$}\footnote{Based on data obtained at the W. M. Keck Observatory, which is operated as a scientific partnership among the California Institute of Technology, the University of California and the National Aeronautics and Space Administration. The Observatory was made possible by the generous financial support of the W. M. Keck Foundation.}}

\correspondingauthor{Zahra Sattari}
\email{zahra.sattari@email.ucr.edu}

\author{Zahra Sattari}
\affiliation{Department of Physics and Astronomy, University of California, Riverside, 900 University Ave, Riverside, CA 92521, USA}

\author{Bahram Mobasher}
\affiliation{Department of Physics and Astronomy, University of California, Riverside, 900 University Ave, Riverside, CA 92521, USA}

\author{Nima Chartab}
\affiliation{Department of Physics and Astronomy, University of California, Riverside, 900 University Ave, Riverside, CA 92521, USA}

\author{Behnam Darvish}
\affiliation{Cahill Center for Astrophysics, California Institute of Technology, 1216 East California Boulevard, Pasadena, CA 91125, USA}

\author{Irene Shivaei}
\altaffiliation{Hubble Fellow}
\affiliation{Steward Observatory, University of Arizona, 933 North Cherry Ave, Tucson, AZ 85721, USA}

\author{Nick Scoville}
\affiliation{Cahill Center for Astrophysics, California Institute of Technology, 1216 East California Boulevard, Pasadena, CA 91125, USA}

\author{David Sobral}
\affiliation{Department of Physics, Lancaster University, Lancaster, LA1 4YB, UK}

\begin{abstract}
We study the mass-metallicity relation for 19 members of a spectroscopically-confirmed protocluster in the COSMOS field at $z=2.2$ (CC2.2), and compare it with that of 24 similarly selected field galaxies at the same redshift. Both samples are $\rm H\alpha$ emitting sources, chosen from the HiZELS narrow-band survey, with metallicities derived from $\rm N2\ (\frac{\rm [NII] \lambda 6584}{\rm H \alpha})$ line ratio. For the mass-matched samples of protocluster and field galaxies, we find that protocluster galaxies with $10^{9.9} \rm M_\odot \leq M_* \leq 10^{10.9} \rm M_\odot$ are metal deficient by $0.10 \pm 0.04$ dex ($2.5\sigma$ significance) compared to their coeval field galaxies. This metal deficiency is absent for low mass galaxies, $\rm M_* < 10^{9.9} \rm M_\odot$. Moreover, relying on both SED-derived and $\rm {H\alpha}$ (corrected for dust extinction based on $\rm {M_*}$) SFRs, we find no strong environmental dependence of SFR-$\rm {M_*}$ relation, however, we are not able to rule out the existence of small dependence due to inherent uncertainties in both SFR estimators. The existence of $2.5\sigma$ significant metal deficiency for massive protocluster galaxies favors a model in which funneling of the primordial cold gas through filaments dilutes the metal content of protoclusters at high redshifts ($z \gtrsim 2$). At these redshifts, gas reservoirs in filaments are dense enough to cool down rapidly and fall into the potential well of the protocluster to lower the gas-phase metallicity of galaxies. Moreover, part of this metal deficiency could be originated from galaxy interactions which are more prevalent in dense environments.   

\end{abstract}

\section{Introduction}
In the standard $\rm \Lambda CDM$ cosmological scenario, structures form from the growth of small fluctuations through gravitational instability. Dark matter structures grow by two main processes: By merging small halos to form the larger ones or by smooth accretion of dark matter from their immediate environment. Baryons fall into the potential well of these dark matter structures and feed galaxies within those structures with cold pristine gas, which allows them to form their stars. Galaxy clusters observed in the present Universe are the largest virialized dark matter halos that are populated with massive and evolved galaxies \citep[e.g.,][]{Dressler80,Balogh04,Kauffmann04,Peng10}. At high redshifts ($z \gtrsim 2$), most of these clusters have not yet had the time to virialize and consist of subhalos which will then merge and form the massive clusters present in the local Universe. Direct observations of the progenitors of these clusters at high redshifts, known as protoclusters, provide useful information about the processes involved in the early stages of structure formation. 

It is now well established that, at a given redshift, the interstellar medium (ISM) of massive galaxies is more metal-enriched than that of low mass galaxies. This correlation, known as mass-metallicity relation (hereafter MZR), is observed out to $z \sim 3.5$ \citep[see review by][and references therein]{Maiolino19}. The infall of cold gas from intergalactic medium (IGM) provides the fuel for galaxies to form their stars, which are the factories responsible for metal production in galaxies. However, galaxies are not very efficient in forming stars as the cold gas accretes into their potential well. \cite{Behroozi13} found that the star formation efficiency of galaxies (the star formation rate (SFR) divided by the baryon accretion rate) can reach a maximum of $\sim 55 \%$. Therefore, the extra primordial cold gas that could not convert to stars dilutes the metal content of the ISM. This mechanism can explain observations where, at a given stellar mass, galaxies with higher SFR have lower gas-phase metallicity \citep[e.g.,][]{Mannucci10,Lara10,Stott13,Sanders2018}. More interestingly, many studies suggest that these two parameters, SFR and gas accretion rate, depend on the environment of galaxies even at high redshift Universe \citep{Darvish16,Kawinwanichakij17,Chartab20}. Thus, comparing gas-phase metallicity of galaxies in extreme environments and high redshifts, such as protoclusters, with a similarly selected population of field galaxies at the same redshift can provide valuable insights about galaxy evolution processes and their environmental dependencies.

Although in the local Universe, most of the studies found evidence for environmental imprint on the MZR \citep[e.g.,][]{Cooper08,Ellison09,Peng14,Sobral15}, environmental dependence of the MZR at high redshift is still controversial. Some studies found an enhancement in the gas-phase metallicity of galaxies in protoclusters compared to their field counterparts at $z\sim 2$ \citep{Kulas13,Shimakawa15}. However, there is observational evidence of metal deficiency in protocluster galaxies at $z \sim 2$ compared to their field counterparts, according to \cite{Valentino15}. Moreover, \cite{Chartab21} studied the relation between the local density of MOSFIRE Deep Evolution Field \citep[MOSDEF;][]{Kriek15} galaxies and their gas-phase metallicities and found that galaxies in overdensities have $\sim 0.07$ dex lower metallicity than the field galaxies at $z\sim 2.3$. On the other hand, \cite{Kacprzak15} and \cite{Alcorn19} found no significant environmental dependence on the MZR of galaxies at $z \sim 2$. 

In this paper, we study the MZR for galaxy members of the recently confirmed protocluster, CC2.2, in the Cosmic Evolution Survey \citep[COSMOS;][]{Scoville07} at $z \sim 2.2$ \citep{Darvish20}.  We then compare this relation with a control sample of similarly selected field galaxies. Both protocluster and field samples are $\rm H\alpha$ emitting sources, selected from the narrow-band High-Z Emission Line Survey \citep[HiZELS;][]{Geach12, Sobral13, Sobral14}. The paper is organized as follows: In Section \ref{Data}, we describe the KeckI/MOSFIRE observations and data reduction procedure followed by stellar mass measurements and sample selection. We then explain the stacking process and measurements of gas-phase metallicities in Section \ref{Measurements}. In Section \ref{Results}, we construct the MZR for both protocluster and field samples and compare them to deduce the role of the environment in MZR at $z \sim 2.2$. We discuss our results in Section \ref{Discussion}.

Throughout this paper, we assume a flat $\Lambda$CDM cosmology with $H_0=70 \rm \ kms^{-1} Mpc^{-1}$, $\Omega_{m_{0}}=0.3$ and $\Omega_{\Lambda_{0}}=0.7$. All the physical parameters are measured assuming a \cite{Chabrier03} initial mass function (IMF).

\section{Data}\label{Data}

\subsection{MOSFIRE Observation}\label{MOSFIRE}
In this paper, we use near-IR spectroscopy of galaxies in a recently confirmed protocluster, CC2.2, at $z \sim 2.2$ \citep{Darvish20}, in the COSMOS field \citep{Scoville07}. The spectroscopic observations were conducted with KeckI/MOSFIRE NIR multi-object spectrograph \citep{McLean12} in December 2018 and January 2019 in both $K(\sim 1.92-2.40 \mu m)$ and $H (\sim 1.47-1.81 \mu m)$ bands, leading to 35 confirmed members. The primary spectroscopic sample comprises the narrow-band $\rm H \alpha$ emitting candidates from the HiZELS survey \citep{Geach12, Sobral13, Sobral14}. For a full description of the protocluster identification and observation, we refer readers to \cite{Darvish20}.  

As a control sample, we use KeckI/MOSFIRE spectroscopic observations of 24 field galaxies in the redshift range $2.22 \leq z \leq 2.24$, located in less-crowded regions of the COSMOS (16 galaxies) and UDS (8 galaxies) fields. All these field galaxies were observed over the observing programs during 2018-2019 (PI: N. Z. Scoville) in $K$ band (a few in $H$ band as well) and are selected similarly from the HiZELS survey to avoid any biases introduced by sample selection.

The observations were performed the same way for both field and protocluster galaxies, minimizing any source of bias. All the observing nights were conducted under clear conditions with the average seeing of $0.5^{\prime \prime}$, and a typical exposure time of $\sim 90$ minutes per mask.

\begin{figure*}[]
    \centering
	\includegraphics[width=1\textwidth,clip=True, trim=6cm 0.75cm 6cm 0.75cm]{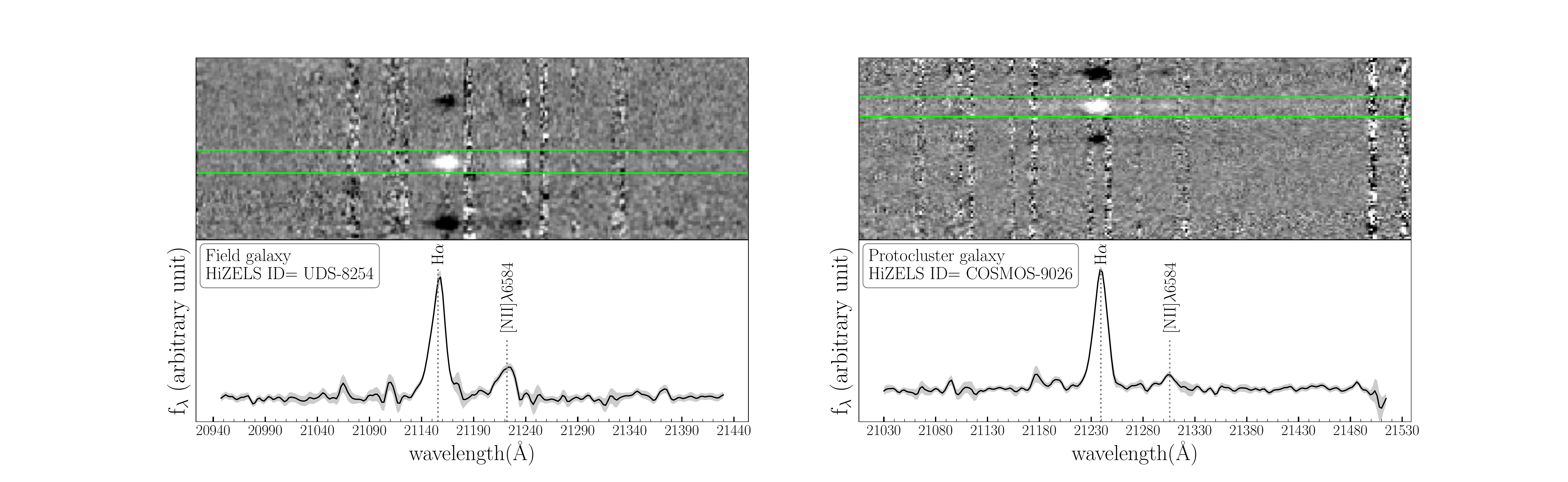}
	\caption{The 2D and corresponding optimally extracted 1D spectra for two galaxies selected from our sample. \textit{Left}: Example of a field galaxy. This galaxy is at $z=2.223$ and has a stellar mass $10^{10.05} \rm M_{\odot}$. \textit{Right}: A galaxy member of the protocluster at $z=2.235$. The stellar mass of this galaxy is $10^{9.96} \rm M_{\odot}$. In both spectra, $\rm H \alpha$ and $\rm [NII] \lambda 6584$ emission lines are evident.}
	\label{fig:1Dspectrum} 
\end{figure*}

\subsection{Data Reduction}\label{data_reduction}
The acquired data were reduced using the MOSFIRE data reduction pipeline (DRP) \footnote{\hyperlink{https://github.com/Keck-DataReductionPipelines/MosfireDRP}{https://github.com/Keck-DataReductionPipelines/MosfireDRP}}. The outputs of the DRP are rectified, sky-subtracted, and wavelength-calibrated 2D spectra and their associated uncertainties. We then extract optimally weighted 1D spectra and their errors using the optimal extraction algorithm \citep{Horne86}. We weight the flux of each pixel by the inverse of the flux variance and the spatial extent of the 2D spectrum in an optimized window and then, sum the weighted fluxes along the wavelength axis. The size of the optimized window is determined such that the bright features with the highest signal-to-noise (S/N) in the 2D spectrum are surrounded by the window. The weighted summation within the optimized window produces 1D spectra of the sources along with their errors. Figure \ref{fig:1Dspectrum} shows an example of the 2D and optimally extracted 1D spectra for a field galaxy and a protocluster member.

We fit a triple Gaussian function to the reduced 1D spectra to extract $\rm [NII] \lambda 6548$, $\rm H \alpha$ and $\rm [NII] \lambda 6584$ emission line fluxes. We require a constant value for the continuum and the same width for all three emission lines. The line fluxes are the integration of the best fit Gaussian function to the 1D spectra. For the error calculation, we perturb the 1D spectra 1000 times and remeasure the line fluxes. The standard deviation of these 1000 measurements is assigned as the uncertainty in line fluxes. The redshifts of the sources are measured based on the peak of $\rm H \alpha$ emission lines with $\rm S/N \geq 3$.

\subsection{Stellar Masses and SFRs}\label{mass}
We estimate the stellar masses ($\rm M_*$) and SFRs of galaxies by fitting synthetic spectral energy distributions (SED) to their available photometric data (COSMOS: \cite{Laigle16}; UDS: \cite{Mehta18}). To perform the SED fitting, we utilize Bayesian Analysis of Galaxies for Physical Inference and Parameter EStimation (Bagpipes) code \citep{Carnall18}, which uses 2016 version of a library of \cite{Bruzual03} synthetic spectra. We fix the redshifts of galaxies to their spectroscopic values, considering delayed exponentially declining star formation history, $\rm t e^{-t/\tau}$. A range of $0.3-10$ Gyr with a uniform prior is assumed for the star formation e-folding time-scale ($\tau$). The nebular emission models which are constructed based on the methodology of \cite{Byler17} are added to the SEDs. Also a metallicity range $0<\rm Z/Z_\odot<2.5$ with a logarithmic prior and a \cite{Calzetti00} extinction law are adopted.

The SFRs of the galaxies are also estimated using their $\rm H \alpha$ emission line fluxes taken from HiZELS survey \citep{Sobral13}. Since the $H$ band data are not available for most of the sample, following \cite{Sobral12} and \cite{Koyama13}, we utilize \cite{Garn10} calibration to correct the $\rm {H \alpha}$ luminosity for the dust extinction based on the stellar masses of the galaxies. We then convert the dust-corrected $\rm {H \alpha}$ luminosity to SFR using the calibration from \cite{Kennicutt98}: $\rm SFR(M_\odot yr^{-1})=7.9 \times 10^{-42} L_{H \alpha}(erg/s).$ One should note that the SFR derived from this method is highly uncertain due to the existence of a large scatter in dust attenuation calibration based on the stellar mass.

\subsection{Sample Selection}
We select galaxies with significant detection in $\rm H \alpha$ emission lines (S/N$\geq 3$). We exclude galaxies with $\rm M_* < 10^{9.5} \rm M_{\odot}$ to construct a mass complete sample. We also remove potential mergers by visual inspection of their spectra and images, as well as the active galactic nuclei (AGNs). The AGNs are identified through their broad emission lines, X-ray flags included in the UDS and COSMOS catalogs, or IR emissions \citep{Donley12}. Moreover, optical AGNs are excluded by requiring $\log (\frac{\rm [NII] \lambda 6584}{\rm H \alpha})<-0.3$ \citep{Coil15}. These criteria result in 19  protocluster members at $z=2.23$ and 24 field galaxies at the same redshift, spanning a narrow redshift range $2.22 \leq z \leq 2.24$.

\section{Composite spectra}\label{Measurements}

\begin{figure*}
    \centering
	\includegraphics[width=1.1\textwidth,clip=True, trim=2.4cm 0cm 0cm 1cm]{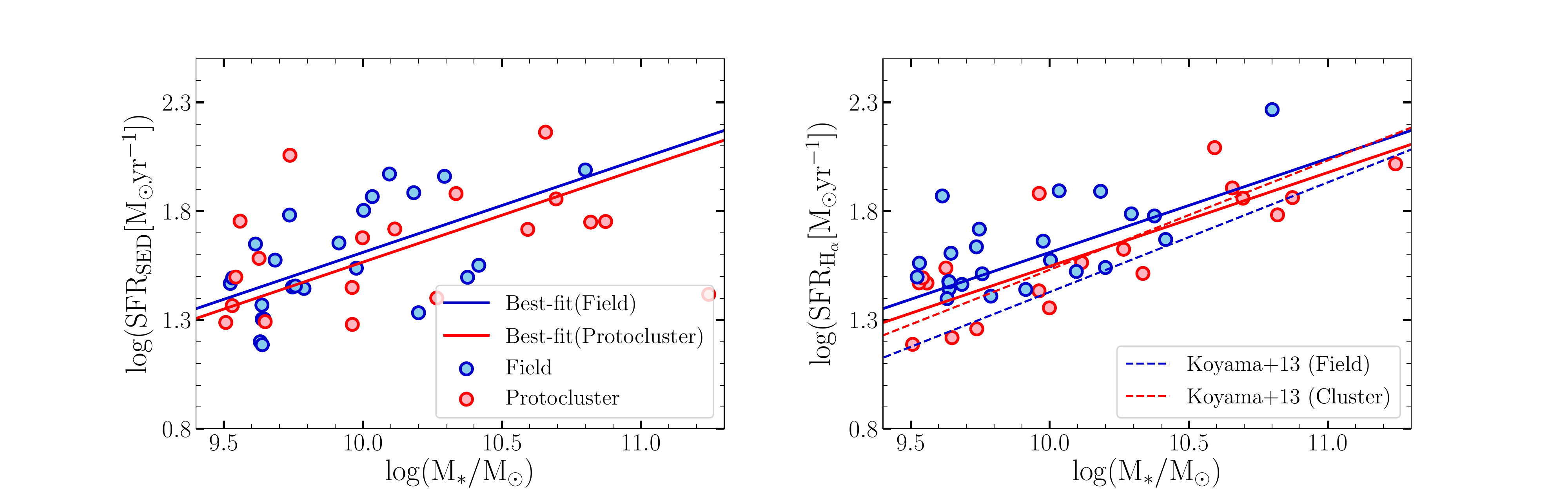}
	\caption{SFR of the field (blue circles) and protocluster (red circles) galaxies as a function of their stellar masses. \textit{Left}: The SFRs are calculated from SED fitting. \textit{Right}: The SFRs are determined using the $\rm {H \alpha}$ luminosity of galaxies corrected for the dust extinction based on their stellar masses. The solid lines in each panel show the best-fit to the data points. The slope of the protocluster best-fit line (solid red lines) is fixed at the slope of the best-fit line for field galaxies (solid blue lines). The dashed blue (red) line shows the best-fit for the field (cluster) galaxies from \cite{Koyama13}.}\label{SFR} 
\end{figure*}

\subsection{Stacking Analysis}\label{stacking}

We measure the gas-phase metallicity of galaxies, using the $\rm N2$ ($\frac{\rm [NII] \lambda 6584}{\rm H \alpha}$) line ratio. Due to the prevalence of $\rm [NII] \lambda 6584$ undetected galaxies in our low-mass sample ($\rm S/N<3$), we use stacking technique. To stack the spectra, we divide the sample into three stellar mass bins, with an equal number of objects in each bin, for both the protocluster and field galaxies. We provide the range of stellar mass bins and the number of galaxies residing in each bin for the field and protocluster samples in Table \ref{mass_bin_range}. We then shift the spectra of galaxies to their rest-frame and normalize them by the total $\rm H \alpha$ luminosity. In each mass-bin, we bin the normalized spectrum with a resolution of $0.5\ \text{\AA}$. The stacked spectrum is calculated as the weighted average of the spectra in each $0.5\ \text{\AA}$ bin:

\begin{equation}
    \Tilde{f}(\lambda)^{\rm{stacked}}=\frac{\displaystyle\sum_{i} \frac{\Tilde{f}_i(\lambda)}{{\sigma_i}^2}}{\displaystyle\sum_{i} \frac{1}{{\sigma_i}^2}},
\end{equation} where $\Tilde{f}_i(\lambda)$ is the flux density of each normalized spectrum, $\sigma_i$ is its corresponding standard deviation, and $\Tilde{f}(\lambda)^{\rm{stacked}}$ is the composite spectrum with the uncertainty of $\sqrt{1/(\displaystyle\sum_{i} \frac{1}{{\sigma_i}^2})}$ in each mass-bin. The protocluster and field galaxies are also bootstrap resampled 100 times to take into account the sample variance. To perform the bootstrap resampling, we draw a random sample of galaxies from the original sample considering replacement. The replacement allows us to have a random sample that may include some duplicate members from the original sample, or may not contain some of the galaxies from the initial sample. This process is repeated 100 times and each time we end up having new stacked spectra. The sample variance is calculated using the standard deviation of these 100 trials.

\begin{table}[!h]
\centering
\caption{The stellar mass range of each mass bin in the stacking process}
\label{mass_bin_range}
\begin{tabular}{|c|c|c|}
\hline
Sample                        & $\rm \log{\frac{M_*}{M_\odot}}$                                            & Number of galaxies \\ \hline
\multirow{3}{*}{Field}        & {[}9.5,9.7)       & 8                  \\ \cline{2-3} 
                              & {[}9.7,10)        & 8                  \\ \cline{2-3} 
                              & {[}10,10.8{]}   & 8                  \\ \hline
\multirow{3}{*}{Protocluster} & {[}9.5,9.8)       & 7                  \\ \cline{2-3} 
                              & {[}9.8,10.5)      & 6                  \\ \cline{2-3} 
                              & {[}10.5,11.2{]} & 6                  \\ \hline
\end{tabular}
\end{table}

\subsection{Metallicities}\label{metallicity}

Since only a few field galaxies are observed in the $H$ band ($\sim10 \%$ of the field galaxies), we utilize rest-frame optical emission line in the observed $K$ band to measure the gas-phase metallicities. This is done by measuring the best-fit $\rm [NII] \lambda 6584$ and $\rm H \alpha$ emission line intensities and estimating the gas-phase metallicity using the $\rm N2\ (\frac{\rm [NII] \lambda 6584}{\rm H \alpha})$ line ratio. 

To measure the line fluxes and their uncertainties for the stacked spectra, we use the same methodology described in Section \ref{data_reduction} for the individual spectra. Since the stacked spectra are normalized by the $\rm H \alpha$ luminosity, the area underneath the $\rm [NII] \lambda 6584$ line in the stacked spectra corresponds to $\langle \frac{\rm [NII] \lambda 6584}{\rm H \alpha}\rangle$.  

In order to measure the gas-phase oxygen abundance ($12+\log (\rm O/H)$) of galaxies, we employ the empirical calibration from \cite{Pettini04}, which is based on electron temperature measurements in the local $\rm H II$ regions, and is given by $\rm 8.9+0.57 \log (N2)$.

\section{Results}\label{Results}
\subsection{SFR-${M_*}$ Relation}
Different studies have shown that the fundamental metallicity relation (FMR) exists for galaxies at $z \sim 2$ \citep{Mannucci10,Sanders2018}. According to this relation, at a given stellar mass, galaxies that have a higher SFR tend to have lower gas-phase metallicities. Thus, before studying the environmental dependence of the MZR in our sample, to investigate any possible dependence between the environment of the galaxies and their SFRs we show the relation between the SFR of the field and protocluster galaxies and their stellar masses in Figure \ref{SFR}. The SFR in the left panel is calculated from SED fitting, while the right panel shows the $\rm {H \alpha}$ SFRs (corrected for extinction as mentioned in section \ref{mass}) as a function of stellar mass.

An important issue in the SFR estimation using these two methods is that there are large uncertainties in both measurements. This will not allow us to rule out any small environmental dependence of the main sequence, if exists. For a better comparison, we also add the best-fit line from \cite{Koyama13} in the right panel of Figure \ref{SFR}. Our result is in agreement with theirs, showing no significant environmental dependence of the main sequence galaxies for our star forming sample.

\subsection{Mass-Metallicity Relation}\label{MZR}

\begin{figure*}
    \centering
	\includegraphics[width=0.8\textwidth,clip=True, trim=0.72cm 0.75cm 0.75cm 0.75cm]{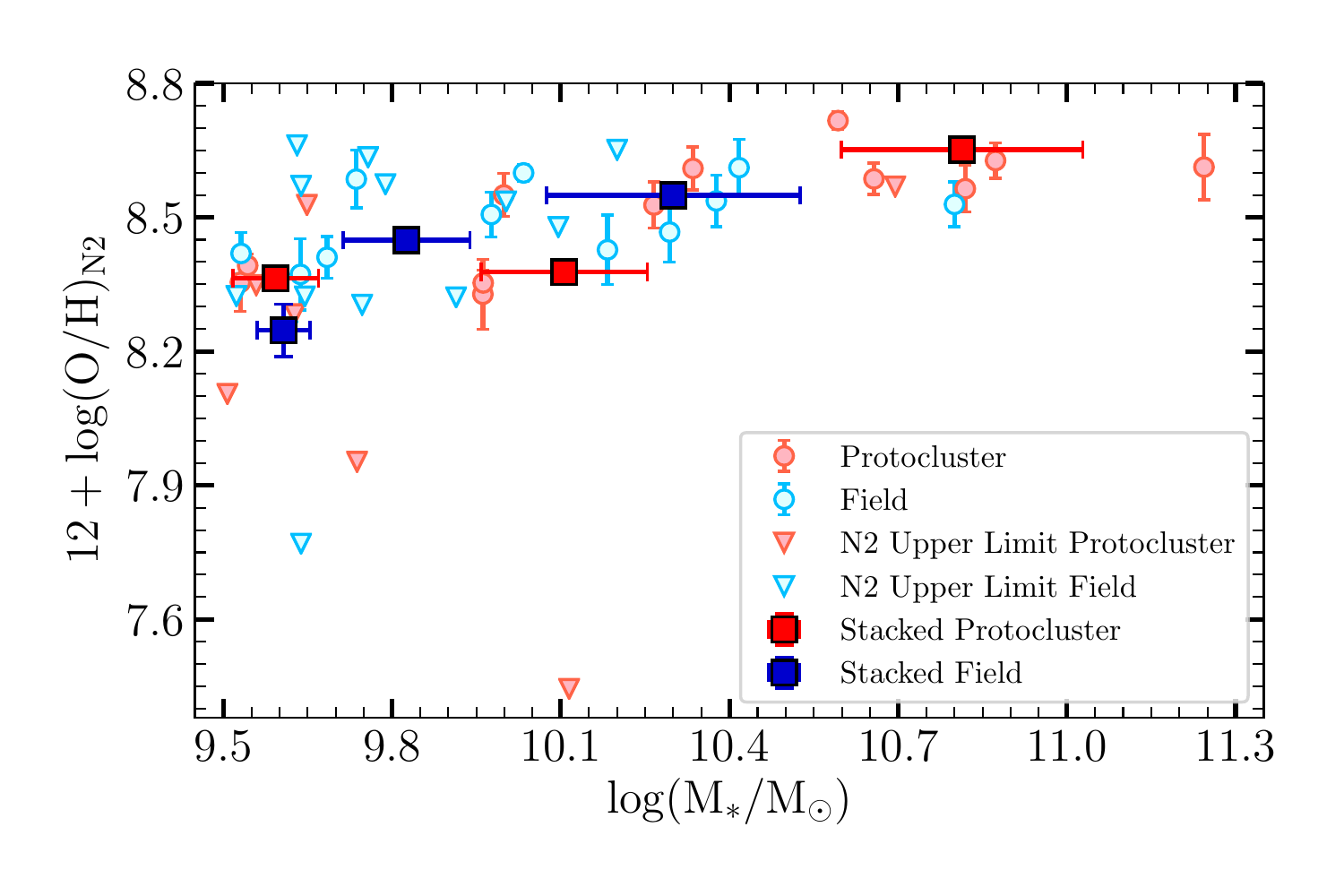}
	\caption{MZR for the field (blue circles) and protocluster (red circles) galaxies at $z \sim 2.2$, without controlling for the stellar mass distribution (mass-matching). The $1 \sigma$ upper-limits for galaxies with undetected $\rm [NII] \lambda 6584$ line are shown with inverted triangles. The solid blue (red) squares indicate the metallicity measurements for stacked spectra of field (protocluster) galaxies in three bins of stellar mass. The error bars are smaller than the square symbols if not shown. Also, the horizontal error bars in the stacked data show the $1 \sigma$ scatter in the stellar mass of each bin.}\label{MZR_NMM} 
\end{figure*}

In this section, we study the MZR for the field and protocluster samples to investigate the role of the environment in the gas-phase metallicity of galaxies at fixed $\rm M_*$. As discussed in Section \ref{stacking}, we divide the sample into three stellar mass bins with equal number of galaxies in each bin. Figure \ref{MZR_NMM} shows the MZR for the stacked spectra and individual galaxies. The metallicities of the stacked spectra in three stellar mass bins are shown in blue (red) squares for field (protocluster) galaxies. The stellar mass uncertainty in the stacked data points shows $1 \sigma$ scatter in each bin.

As shown in Figure \ref{MZR_NMM}, in the stellar mass range $10^{9.7} \rm M_{\odot} \lesssim \rm M_* \lesssim 10^{10.5} \rm M_{\odot}$, the average protocluster galaxies have a relatively lower metallicity than the average field galaxies by $\sim 0.1$ dex. However, the gas-phase metallicity of low-mass galaxies, $\rm M_* < 10^{9.7} \rm M_{\odot}$, do not significantly depend on the environment. Moreover, in the massive end of the MZR ($\rm M_* > 10^{10.5} \rm M_{\odot}$), due to the small number of field galaxies, we cannot draw robust conclusions on the MZR variation between the field and protocluster galaxies. In the following section, we match the stellar mass distributions of protocluster and field samples to properly isolate the effect of stellar mass from galaxy environment.

\subsection{The Mass-Matched Samples}
It is known that protoclusters often host more massive galaxies compared to the field. Therefore, to have a reliable comparison between the metallicity of protocluster and field galaxies at fixed stellar mass, the two samples should have similar stellar mass distributions. Otherwise, any change in the MZR may be attributed to the differences in stellar mass distributions. 

\begin{figure*}[!t]
    \centering
	\includegraphics[width=1\textwidth,clip=True, trim=1cm 0cm 0.5cm 0cm]{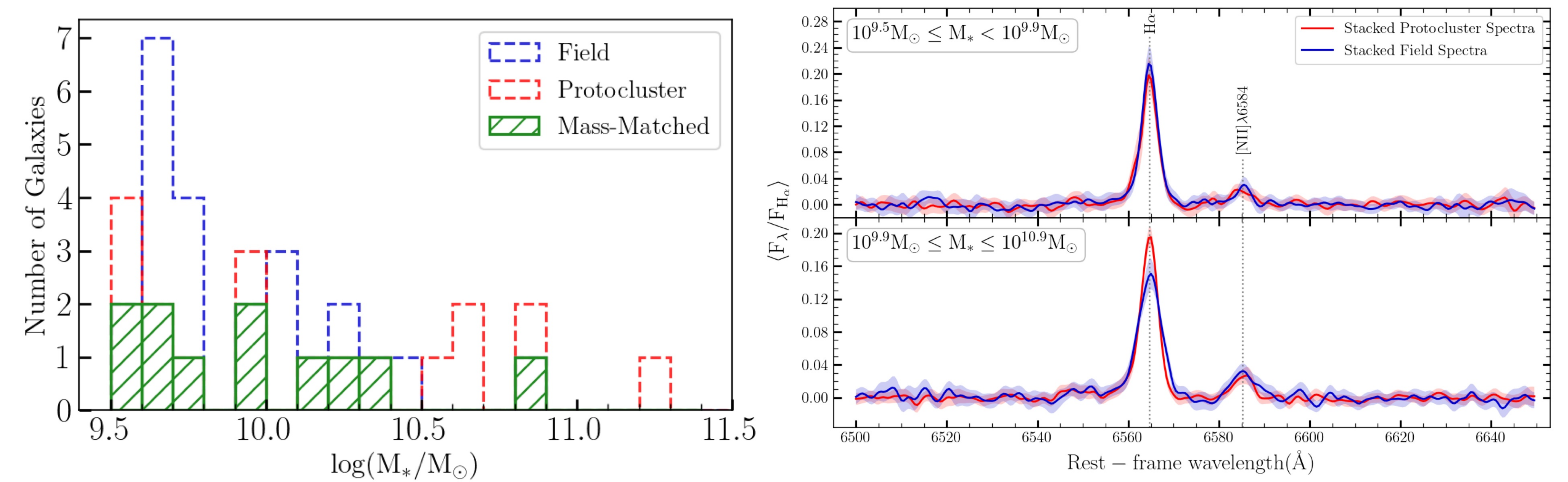}
	\caption{\textit{Left}: The stellar mass distributions for the field (blue) and protocluster (red) galaxies. The green hatched region shows the matched stellar mass distributions of protocluster and field galaxies. \textit{Right}: Composite spectra for the mass-matched samples in two bins of stellar mass ($10^{9.5} \rm M_\odot \leq M_* < 10^{9.9} \rm M_\odot$, and $10^{9.9} \rm M_\odot \leq M_* \leq 10^{10.9} \rm M_\odot$). The stacked spectra of protocluster galaxies are shown in red and the stacked spectra of field galaxies are shown in blue.}
	\label{hist_stack} 
\end{figure*}

The left panel in Figure \ref{hist_stack} shows the stellar mass distribution for the field and protocluster galaxies. It is clear that, for protocluster galaxies in the massive end of the distribution, there is no analog of the field galaxy, resulting in a biased comparison between field and protocluster samples. We resolve this by constructing the mass-matched samples. Similar to \cite{Chartab21}, we match the stellar mass distributions of field and protocluster galaxies with the resolution of $\rm \log(M_*/M_\odot)=0.1$ dex, which is the typical stellar mass uncertainties computed in section \ref{mass}. As a result of mass-matched distributions, we have the same number of protocluster and field galaxies in each stellar mass bin of $\rm \log(M_*/M_\odot)=0.1$ (green hatched region in the left panel of Figure \ref{hist_stack}).

However, there is not a unique way to subsample data and construct mass-matched samples. For instance, four protocluster galaxies are in the mass range $10^{9.5} \rm  M_\odot \leq M_* \leq 10^{9.6} \rm M_\odot$, but just two field galaxies are in this range of stellar mass. Thus, to take into account all the galaxies that reside in each mass-matched region, we randomly subsample galaxies 500 times and each time, we construct the stacked spectra and perturb them based on their uncertainties. The stacking process is the same as described in Section \ref{stacking}, where we consider both measurement errors and sample variance. The average and standard deviation of 500 trials correspond to the composite spectrum and its error, respectively. The right panels in Figure \ref{hist_stack} show the composite spectra for the mass-matched sample of protocluster and the field galaxies in two stellar mass bins $10^{9.5} \rm M_\odot \leq M_* < 10^{9.9} \rm M_\odot$, and $10^{9.9} \rm M_\odot \leq M_* \leq 10^{10.9} \rm M_\odot$.

We report the gas-phase metallicities of the field and protocluster galaxies for the mass-matched samples in Table \ref{metallicities}. Red (blue) squares in Figure \ref{massmatched-compare} (top panel) show the MZR for the stacked protocluster (field) sample in two stellar mass bins $10^{9.5} \rm M_\odot \leq M_* < 10^{9.9} \rm M_\odot$, and $10^{9.9} \rm M_\odot \leq M_* \leq 10^{10.9} \rm M_\odot$. The metallicity estimates for all the galaxies in the mass-matched samples are also included in the figure. At $z \sim 2.2$, the protocluster galaxies in the massive end of the MZR are metal deficient by $0.10 \pm 0.04$ $(2.5\sigma$ significance) dex compared to those residing in the field. However, this deficiency is not significant ($< 1\ \sigma$) in the lower mass bin ($0.03 \pm 0.06$ dex), possibly due to the prevalence of non-detections and/or small sample size.

\begin{table}[!h]
\caption{Gas-phase metallicities of the stacked spectra for the mass-matched samples. The second and third columns show stellar mass ranges and the mean value of mass in each range, respectively.}
\label{metallicities}
\begin{tabular}{|c|c|c|c|}
\hline
Environment                   & \multicolumn{1}{l|}{$\rm \log{\frac{M_*}{M_\odot}}$} & $\rm \langle \log{\frac{M_*}{M_\odot}}\rangle$ & $\rm \langle 12+\log(O/H) \rangle$ \\ \hline
\multirow{2}{*}{Field}        & {[}9.5,9.9)                                           & $9.62 \pm 0.04$                                & $8.38 \pm 0.04$                    \\ \cline{2-4} 
                              & {[}9.9,10.9{]}                                        & $10.25 \pm 0.12$                               & $8.50 \pm 0.03$                    \\ \hline
\multirow{2}{*}{Protocluster} & {[}9.5,9.9)                                           & $9.62 \pm 0.03$                                & $8.35 \pm 0.05$                    \\ \cline{2-4} 
                              & {[}9.9,10.9{]}                                        & $10.25 \pm 0.12$                               & $8.40 \pm 0.03$                    \\ \hline
\end{tabular}
\end{table}

\subsection{Comparison with Literature}

\begin{figure*}%
    \centering
    \subfloat{{\includegraphics[width=0.8\textwidth,clip=True, trim=0cm 0.73cm 0cm 0.6cm]{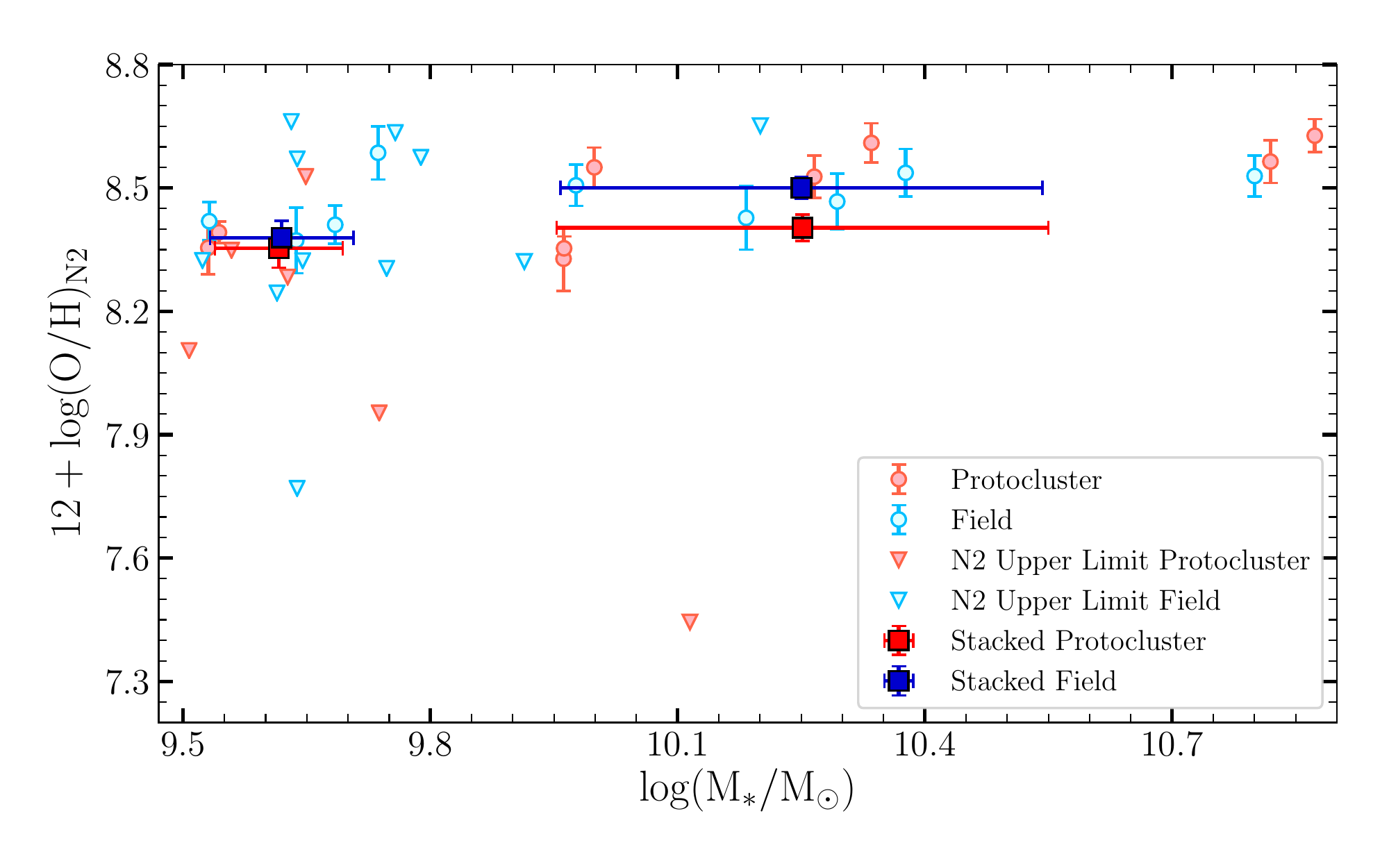}}}%
    \qquad
    \subfloat{{\includegraphics[width=0.82\textwidth,clip=True, trim=0cm 0.73cm 0cm 0.6cm]{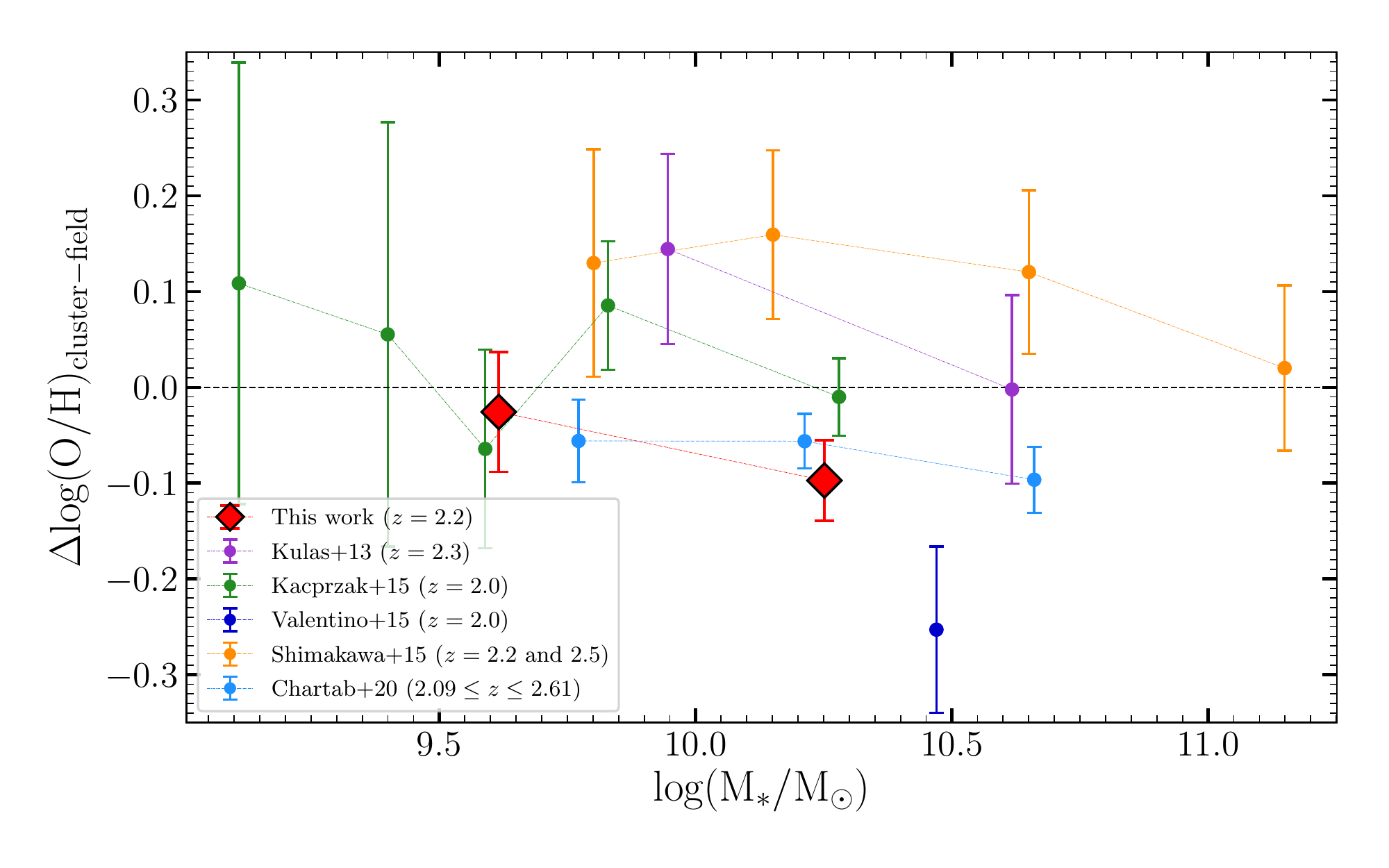}}}%
    \caption{\textit{Top}: The MZR for the mass-matched samples at $z \sim 2.2$ for protocluster galaxies (red) and field galaxies (blue). Inverted triangles show $\rm [NII] \lambda 6584$ non-detections. The gas-phase metallicities for stacked spectra of protocluster and field samples in two stellar mass bins of $10^{9.5} \rm M_\odot \leq M_* < 10^{9.9} \rm M_\odot$, and $10^{9.9} \rm M_\odot \leq M_* \leq 10^{10.9} \rm M_\odot$ are also shown in red and blue squares, respectively. The horizontal error bars in the stacked data show $1 \sigma$ scatter in stellar mass of each bin. \textit{Bottom}: The compilation of the difference (offset) between gas-phase metallicity of protocluster and field galaxies as a function of stellar mass from literature. The red data points show the offset for the mass-matched samples used in this work in two stellar mass bins. The uncertainties in the offsets include errors from gas-phase metallicities of both protocluster and field samples.}%
    \label{massmatched-compare}%
\end{figure*}

To compare our results with those in the literature, we show the offset between the average gas-phase metallicity of the protocluster (galaxies in overdense regions) and the field galaxies from different studies (including the present work) as a function of stellar mass in Figure \ref{massmatched-compare} (bottom panel).

We emphasize that gas-phase metallicity is calibrated locally and its absolute value at high redshift could be uncertain \citep{Steidel14,Shapley19}. However, when we study relative metallicity and estimate the difference of metallicities between field and protocluster galaxies, the calibration effect is not a concern.

Moreover, consideration of selection biases is essential in measuring the MZR \citep{Stott13}. Different selection criteria for protocluster and field samples result in an unreliable comparison between their respective metallicities. As both protocluster and field samples in the present work are $\rm H\alpha$-selected, the metallicity offset does not suffer from such selection biases.    

\cite{Kulas13} studied the MZR of a protocluster sample at $z=2.3$ and compared it with field galaxies at the same redshift. Both samples are selected based on their rest-frame UV emission. The gas-phase metallicity of their sources is calculated using $\rm N2$ indicator. In the lower stellar mass bin ($\rm M_* \sim 10^{10} \rm M_\odot$), they found an offset of $0.15$ dex between the metallicity of protocluster and field galaxies, i.e., their field sample is more metal deficient than the protocluster. Our result is in contrast with their findings, possibly due to the fact that they did not employ mass-matched samples.

Also, in the stellar mass range covered in this paper, \cite{Shimakawa15} found metallicity enhancement ($\sim 0.15$ dex) for two protoclusters at $z=2.2$ and $z=2.5$ compared to the field sample of \cite{Erb06} at $z=2.2$. A part of this enhancement can be caused by different selection criteria they used for protocluster and field samples (Their protoclusters are narrow-band selected, but the field sample from \cite{Erb06} is UV-selected). Comparing our field sample with the UV-selected sample of \cite{Erb06}, we notice that in the low-mass end of the MZR, the narrow-band selected sample has systematically higher metallicity compared to the UV-selected sample. Thus, the disagreement between the present work and \cite{Shimakawa15} can be originated from selection biases.

\cite{Kacprzak15} found no significant difference between the MZR of a protocluster at $z=2$ and a field sample at the same redshift. On the other hand, \cite{Valentino15} found that a protocluster sample at $z \sim 2$ with $\rm M_* \sim 10^{10.5} \rm M_\odot$ is $0.25$ dex metal deficient compared to field galaxies at the same redshift. Their results are in qualitative agreement with the results in this study; however, we find $\sim 0.1$ dex metal deficiency for protocluster galaxies compared to the field sample in the massive end of the MZR at $z\sim 2.2$. Moreover, \cite{Chartab21} recently studied the environmental dependence of the MZR for a sample of H-band selected galaxies in the MOSDEF survey at $1.37 \leq z \leq 2.61$. For a mass-matched sample in the redshift range $2.09 \leq z \leq 2.61$, they found $\sim 0.07$ dex metal deficiency for galaxies in overdense regions compared to field galaxies. As shown in the bottom panel of Figure \ref{massmatched-compare}, our results are in agreement with their findings. Additionally, they found that this metal deficiency increases by the stellar mass, which is also seen in our result in Figure \ref{massmatched-compare} (top panel).

\section{Summary and Discussion}\label{Discussion}
In this paper, we studied the mass-metallicity relation for 19 galaxies in a spectroscopically confirmed protocluster at $z \sim 2.2$ in the COSMOS field and compared it with the MZR of a field sample with 24 galaxies at the same redshift. We used $\frac{\rm [NII] \lambda 6584}{\rm H \alpha}$ ratio to measure the gas-phase metallicity of these galaxies. After matching the stellar mass distributions of field and protocluster samples, we found that the protocluster galaxies with $10^{9.9} \rm M_\odot \leq M_* \leq 10^{10.9} \rm M_\odot$ are $0.10 \pm 0.04$ dex $(2.5\sigma$ significance) metal deficient in comparison to field galaxies at the same redshift. However, this metal deficiency is not significant for low-mass galaxies.

\cite{Darvish20} predicted that this protocluster will grow to a Coma-type cluster with $\sim 9 \times 10^{14} \rm M_\odot$ at $z=0$. \cite{Dekel06} found that at $z \lesssim 2$, halos with  $\rm M_{halo} \gtrsim 10^{12} \rm M_\odot$ will be dominated with hot-mode accretion. However, as we go to higher redshifts, cold streams can penetrate massive halos from filaments hosting dense pristine gas \citep{Keres05}. Based on halo mass evolution trajectories of \cite{Behroozi2013}, we estimate that the progenitor of this protocluster at $z\gtrsim2.5$ has $\rm M_{halo} \lesssim 10^{13.5} \rm M_\odot$, where the protocluster is in a phase that hosts cold streams in hot media \citep{Dekel06}. Therefore, the observed protocluster at $z=2.2$ was experiencing cold streams until $\sim 0.5$ Gyr ago ($z\geq 2.5$), which dilutes the gas-phase metallicity of galaxies residing in the protocluster. After the termination of cold streams, the protocluster galaxies continue to process their gas reservoirs until their star formation fully shuts down. However, $0.5$ Gyr is a short time for the protocluster galaxies to significantly enrich their ISM metal content due to star formation. As a result, we are still witnessing metal deficiency for the members of this protocluster at $z=2.2$ compared to field galaxies. We also expect that the SFRs of the protocluster galaxies increase due to a higher fraction of cold gas, but we are not able to confirm this effect, possibly due to the small sample size and inherent uncertainties in SED-derived and $\rm {H\alpha}$ (mass-dependant extinction corrected) SFRs, which prevent us from detecting weak environmental dependence of SFR at a given $\rm {M_*}$. Future $H$-band spectroscopies of our sample can properly constrain the environmental dependence of SFR given the dust-corrected $\rm {H\alpha}$ luminosities, where the attenuation is derived from Balmer decrements.

In addition, the prevalence of minor/major mergers in dense environments at high redshifts \citep{Hine16,Watson19} can explain a part of the metal deficiency observed in the present work. Although we exclude potential ongoing mergers in our sample, our protocluster galaxies could be descendants of recently-merged galaxies at higher redshifts. Minor and major mergers could provide a higher fraction of cold pristine gas for galaxies, increasing their SFR and lowering their metallicities \citep{Horstman2020}.

Moreover, the stellar mass dependence of metal deficiency (i.e., the absence of significant metal deficiency for low mass galaxies) can be explained by: 1) Massive galaxies located in deeper potential wells, being fed by metal-poor cold streams from the cosmic web \citep{Dekel09}, 2) Scaling of the merger fraction with the stellar mass of the galaxies. \cite{Duncan19} showed that the major merger fraction for massive galaxies ($\rm {M_*>10^{10.3}} \rm {M_{\odot}}$) is 3 times higher than low mass galaxies with $\rm {10^{9.7}} \rm {M_{\odot}<M_*<10^{10.3}} \rm {M_{\odot}}$ at ${z=2}$.

We speculate that the metal content of the protocluster members will rapidly increase at lower redshift as all the remaining cold gas reservoirs will be processed through star formation activity. In addition, the IGM gas within the protocluster will be enriched due to strong outflows, which will then reaccrete into galaxies and enhance their gas-phase metallicity. This metal enhancement at lower redshifts in dense environments compared to field galaxies is observed in previous studies \citep[e.g.,][]{Cooper08,Ellison09,Darvish15}.

\section*{Acknowledgement}
We thank the anonymous referee for providing insightful comments and suggestions that improved the quality of this work. IS is supported by NASA through the NASA Hubble Fellowship grant \# HST-HF2-51420, awarded by the Space Telescope Science Institute, which is operated by the Association of Universities for Research in Astronomy, Inc., for NASA, under contract NAS5-26555. The authors wish to recognize and acknowledge the very significant cultural role and reverence that the summit of Maunakea has always had within the indigenous Hawaiian community. We are most fortunate to have the opportunity to conduct observations from this mountain. This research has been done during the COVID-19 global pandemic. The authors would like to thank all the essential workers who risked their lives allowing us to work from home safely.

\bibliography{MZR}

\end{document}